\documentclass[aps,prb]{revtex4}
\usepackage{graphicx}

\begin{document}
\pagestyle{empty}
\title{Quantum friction and graphene
}
\author{A.I.Volokitin$^{1,2}$\footnote{
\textit{E-mail address}:alevolokitin@yandex.ru}   }
 \affiliation{$^1$Peter Gr\"unberg Institut,
Forschungszentrum J\"ulich, D-52425, Germany} \affiliation{
$^2$Samara State Technical University, 443100 Samara, Russia}

\begin{abstract}

Friction is usually a very complicated process. It appears in its most elementary form when two flat surfaces separated by vacuum gap are sliding relative to each other at zero Kelvin and the friction is generated by the relative movement of quantum fluctuations. For several decades physicists have been intrigued by the idea of quantum friction. It has recently been shown that two non-contacting bodies moving relative to each other experience a friction due to quantum fluctuations inside the bodies \cite{VolokitinRMP2007}. However until recent time there was no experimental evidence for or against this effect, because the predicted friction forces are very small, and precise measurements of quantum forces are incredibly difficult with present
technology. The existence of quantum friction is still debated even among theoreticians \cite{Philbin2009,Pendry2010,VolokitinNJP2011}. However, situation drastically changed with the creation of new material - graphene. We recently proposed \cite{VolokitinPRL2011} that quantum friction can be observed in experiments studying electrical transport phenomena in nonsuspended graphene on amorphous SiO$_2$ substrate.

\end{abstract}

\maketitle \vskip 0.5cm

\section{Introduction}

\begin{figure}
\includegraphics[width=0.30\textwidth]{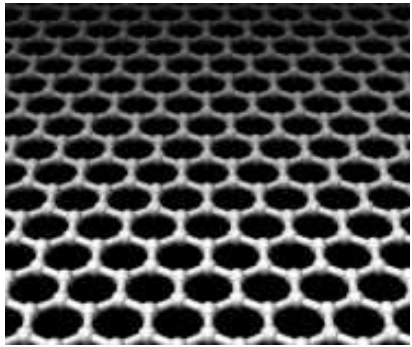}
\caption{\label{graphene} Honeycomb lattice of graphene}
\end{figure}

Graphene, isolated monolayer of carbon, which was learned to obtain very recently, consist of carbon atoms densely packed into two-dimensional honeycomb crystal lattice (Fig. \ref{graphene}). The unique electronic and mechanical properties of graphene are actively studied both theoretically, and experimentally because of their importance for fundamental physics, and also possible technological applications \cite{Geim2004}. In particular, it has been known  theoretically for a long time   that the electron waves in honeycomb crystal lattice can be described by the same equation, as massless fermions in the Dirac relativistic theory. The experimental discovery of graphene made it possible to investigate the phenomena of quantum electrodynamics by studying the electronic properties of this material.

Graphene can also be useful for the detection of quantum friction. The electrons in graphene located on the surface, for example, of the polar dielectric SiO$_2$, will experience additional friction due to interaction with the optical phonons in the dielectric. In  high electric fields the electrons in graphene can move with the very high velocities ($\sim10^6$ m/s). At such velocities the main contribution to the friction will give quantum fluctuations. Thus, quantum friction can be detected by measuring the transport electrical properties of graphene  at  the large electric fields. Besides quantum fluctuations, in the media there are also thermal fluctuations, connected with the thermal motion. They give the contribution to the conservative- dissipative forces of interaction and completely  determine the radiation heat transfer between the  bodies which are at rest. The unusual properties of graphene make it possible to use it for studying the phenomena, connected both with  quantum, and   thermal fluctuations.

\section{Fluctuations produce forces}

There are four known fundamental forces: electromagnetism, gravity, weak and strong interactions. The weak and strong
interactions manifest themselves on length scales on the order of the size of a nucleus, whereas at larger distances
electromagnetism and gravity prevail. It may therefore come as a surprise that two macroscopic non-magnetic bodies with
no net electric charge (or charge moments) can experience an attractive force much stronger than gravity. This force was
predicted by Hendrik Casimir in the late 1940s, and now bears his name. The existence of this force is one of the few direct
macroscopic manifestations of quantum mechanics; others are superfluidity, superconductivity, kaon oscillations, and the black
body radiation spectrum.

\begin{figure}
\includegraphics[width=1.0\textwidth]{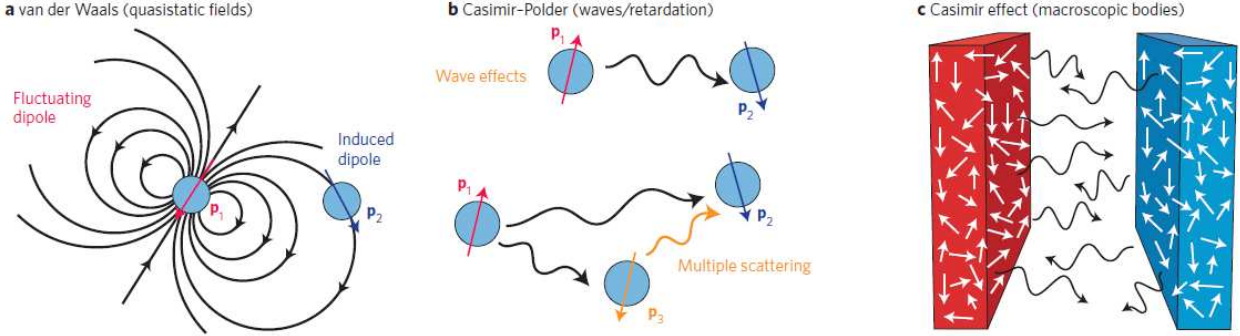}
\caption{\label{CasimirEffect} \textbf{Relationship between van der Waals, Casimir-Polder and Casimir forces, whose origins lie in the quantum fluctuations of dipoles.} \textbf{a}, A fluctuating dipole $\mathbf{p}_1$ induces a fluctuating electromagnetic dipole field, which in turn induces a fluctuating dipole $\mathbf{p}_2$ on a nearby particle, leading to van der Waals forces between the particles. \textbf{b}, When the particle spacing is large, retardation/wave effects modify the interaction, leading to Casimir-Polder forces. When more than two particles interact, the non-additive field interactions lead to a breakdown of the pairwise force laws. \textbf{c}, In situations consisting of macroscopic bodies, the interaction between the many fluctuating dipoles present within the bodies leads to Casimir forces.}
\end{figure}

The origin of both the van der Waals and Casimir forces is connected with the existence of quantum and thermal fluctuations.
 Two neutral particles have fluctuating dipole moments resulting from quantum or thermal effects, which, for a particle separation of $d$, lead to a $d^{-6}$ interaction energy that is commonly used, for example, as a long-range attraction term when describing the interactions between atoms and molecules. Physically, this attraction arises as shown in Fig. \ref{CasimirEffect}a; whenever one
particle acquires a spontaneous dipole moment $\mathbf{p}_1$, the resulting dipole electric field (black lines) polarizes the adjacent particle to produce an induced dipole moment $\mathbf{p}_2\sim  d^{-3}$ . Assuming positive polarizabilities, the direction of the dipole fields means that these two dipoles are oriented so as to attract each other, with an interaction energy that scales as $d^{-6}$. This leads to the van der Waals `dispersion' force. The key to more general considerations of Casimir physics is to understand that this  $d^{-6}$ picture of van der Waals forces makes two crucial approximations that are not always valid: it employs the quasi-static approximation to ignore wave effects, and also ignores multiple scattering if there are more than two particles.
The quasi-static approximation assumes that the dipole moment $\mathbf{p}_1$ polarizes the second particle instantaneously, which is valid if $d$ is much smaller than the typical wavelength of the fluctuating fields. For such separations  the retardation effects are negligible. In this separation region the dispersion
force is usually called the van der Waals force.  However, the finite wave propagation speed of light must be taken into account when $d$ is much larger than the typical wavelength, as shown in Fig. \ref{CasimirEffect}b, and it turns out that the resulting Casimir-Polder interaction energy asymptotically scales as $d^{-7}$ for large $d$. At such separations the dispersion forces are usually called Casimir-Polder (for atom-atom and atom-wall interactions)  forces.
These are both relativistic and quantum-mechanical phenomena first described by Casimir and Polder (1948) \cite{CasimirPolder1948}
and by Casimir (1948) \cite{Casimir1948}, respectively. More generally, the interaction is not a simple power law between these limits, but instead depends on an integral of fluctuations at all frequencies scaled by a frequency-dependent polarizability of the particles.

 The presence of multiple particles further complicates the situation because multiple scattering must be considered (Fig. \ref{CasimirEffect}b). For example, with three particles, the initial dipole $\mathbf{p}_1$ will induce polarizations $\mathbf{p}_2$ and $\mathbf{p}_3$ in the other two particles, but $\mathbf{p}_2$ will create its own field that further modifies $\mathbf{p}_3$, and so on. Thus, the interaction between multiple particles is generally non-additive, and there is no two-body force law that can simply be summed to incorporate all interactions. Multiple scattering is negligible for a sufficiently dilute gas or for weak polarizabilities, but it becomes very significant for interactions between two (or more) solid bodies, which consist of many fluctuating dipole moments that all interact in a complicated way through electromagnetic radiation (Fig. \ref{CasimirEffect}c). When these multiple scattering effects are combined with wave retardation in a complete picture, they yield the Casimir force.

Hendrik Casimir based his prediction on a simplified model involving two parallel perfectly conducting plates separated by a vacuum. Although the Casimir force arises from electromagnetic fluctuations, real photons are not involved. Quantum mechanically, these fluctuations can be described in terms of virtual photons of energy equal to the zero-point energies of the electromagnetic modes of the system. By considering the contribution of the electromagnetic field modes to the zero-point energy  of the parallel plate configuration, Casimir predicted an attractive force between the plates. Because only electromagnetic modes that have nodes on both walls can exist within the cavity, the mode frequencies  depend on the separation between the plates, giving rise to a pressure. The force in this case is attractive because the mode density in free space is larger than that between the plates.
Ideal metals are characterized by perfect reflectivity at all frequencies which means that the absorption wavelength is zero. Thus, the Casimir force is universal and valid at any separation distance. It does not transform to the nonrelativistic London forces at short separations.
Due to the difference in these early theoretical approaches to the description of the dispersion forces, the
van der Waals and Casimir-Polder (Casimir) forces were thought of as two different kinds of force rather than two limiting cases of a single physical phenomenon, as they are presently understood.

\begin{figure}
\includegraphics[width=0.45\textwidth]{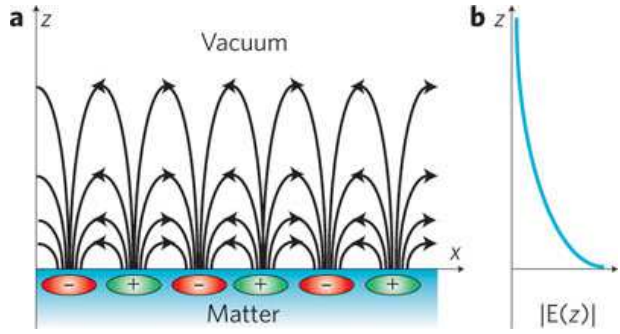}
\caption{\label{evanescent} The amplitude of evanescent electromagnetic waves decays exponentially with the distance away from the surface.
Intensity of the evanescent waves is especially large close to the surface of polar dielectric what is related with the surface phonon polaritons.}
\end{figure}

A unified theory of both the van der Waals and Casimir forces between plane parallel material plates in
thermal equilibrium separated by a vacuum gap was developed by Lifshitz (1955)\cite{Lifshitz1955}. The Lifshitz theory  provides a common tool to deal with dispersive forces in different fields of science (physics, biology, chemistry) and technology.  Lifshitz's theory describes
dispersion forces between dissipative media as a physical phenomenon caused by the fluctuating electromagnetic
field that is always present in both the interior and the  exterior of any medium. Outside the medium this field exists partly in the form of the radiative propagating waves and partly in the form of nonradiative evanescent waves whose amplitudes decay exponentially with the distance away from the medium Fig. \ref{evanescent}. To calculate the fluctuating electromagnetic field Lifshitz used Rytov's theory \cite{Rytov}. Rytov'a theory is based on the introduction into the Maxwell equation of ``random'' field (just as, for example, one introduces a ``random'' force in the theory of Brownian motion). The fundamental characteristic of the random field is the  cross-spectral densities of components of this field at two different points in space. Initially Rytov found this cross-spectral density using phenomenological approach. Later Rytov's formula for the cross-spectral density of the random field was rigorously proved on the base of the fluctuation-dissipation theorem. According to the fluctuation-dissipation theorem, there is a connection between the spectrum of fluctuations of the physical quantity in an equilibrium dissipative medium and the generalized susceptibilities of this medium which describe its reaction to an external influence. Using the theory of the fluctuating electromagnetic field  Lifshitz derived the general formulas for the free energy and force of the dispersion interaction.
In the limit of dilute bodies these formulas describe the dispersion forces acting between atoms and molecules. In the framework of the Lifshitz theory material properties are represented by the frequency dependent dielectric permittivities and atomic properties
by the dynamic atomic polarizabilities. In the limiting cases of small and large separation distances, in
comparison with the characteristic absorption wavelength, the Lifshitz theory reproduces the results obtained
by London and by Casimir and Polder, respectively. It also describes the transition region between the
nonrelativistic and relativistic areas. Both quantum and thermal fluctuations give contributions to the total Casimir force. Quantum fluctuations dominate at small separation ($d\ll \lambda_T=c\hbar/k_BT$) and thermal fluctuations  dominate at large separation ($d\gg \lambda_T$). Casimir forces due to quantum fluctuations have been studied experimentally for a long time. However the Casimir forces due to thermal fluctuations were measured only recently and these measurements confirmed the prediction of the Lifshitz theory \cite{NaturePhysics2011}. At present interest to the Casimit forces increases substantially because it was found that they play exceptionally important role in MEMS and NEMS.

The Lifshitz theory was formulated  for systems at thermal equilibrium. At present there is an interest in the study of
systems out of the thermal equilibrium, in particular in the connection with the
possibility of tuning the strength and sign of the interaction. Such systems also present a way to
explore the role of thermal fluctuations, which usually are masked at thermal equilibrium by the $T=0$ K component, which dominates the
interaction up to very large distances, where the interaction force is very small. In Ref. \cite{Antezza2005} the Casimir-Polder force was
measured at very large distances, and it was shown that the thermal effects on the Casimir-Polder interaction agree with the theoretical prediction. This measurement was done out of thermal equilibrium, where thermal effects are stronger.

Other non-equilibrium thermal effects were investigated by Polder and van Hove \cite{Polder}, who calculated heat flow between two parallel surfaces, divided by vacuum gap. Already more than 30 years  physicists have been interested by  a question about how the Casimir forces and the radiative heat transfer are modified for bodies moving relative to each other. The number of researchers  showed that relative motion of bodies leads to  the friction force  \cite { VolokitinRMP2007}. Theory predicts, that the  friction  force  acts even at   zero temperatures, when it is determined by quantum fluctuations. However, in recent years the existence of quantum friction was   discussed in sufficiently hot debates \cite { Philbin2009, Pendry2010, VolokitinNJP2011}. Philbin and Leonhardt stated \cite {Philbin2009} that all previous theories of quantum friction were incorrect because they included  different approximations. The Philbin-Leonhardt theory  predicts the absence of quantum friction. The general theory of the  Casimir forces and the radiation heat transfer between  moving bodies without  any approximations was developed by us in  Ref. \cite {VolokitinPRB2008}. This theory confirmed correctness of the previous results and showed the inaccuracy of the Philbin-Leonhardt theory  \cite {VolokitinNJP2011}.

\section{Reflection produces friction}

The origin of the van der Waals friction is closely connected with the Casimir forces. The Casimir interaction arises when an atom or
molecule spontaneously develops an electric dipole moment due to  quantum fluctuations. The short-lived atomic polarity can induce a dipole moment in a
neighboring atom or molecule some distance away. The same is true for extended media, where thermal and quantum fluctuation of the current
density in one body induces a current density in other body; the interaction between these current densities is the origin of the Casimir
interaction. When two bodies are in relative motion, the induced current will lag slightly behind the fluctuating current inducing it, and this is
the origin of the van der Waals friction. The Casimir interaction is mostly determined by exchange of the virtual photons between the bodies
(connected with quantum fluctuations), and does not vanish even at zero temperature. Thermal fluctuations begin  noticeably affect Casimir forces only at the large distance between the bodies, when contribution from the quantum fluctuations becomes  very small. On the contrary, van der Waals friction at the low velocities ($v\ll dk_BT/\hbar$) is determined by the exchange of the real photons, connected with the thermal fluctuations. However, at large velocities and low temperatures ($v\gg dk_BT/\hbar$) quantum fluctuations give the main contribution to friction. Specifically, under such conditions quantum friction should be searched, which can be considered as the limiting case of the van der Waals friction.

\begin{figure}
\includegraphics[width=0.45\textwidth]{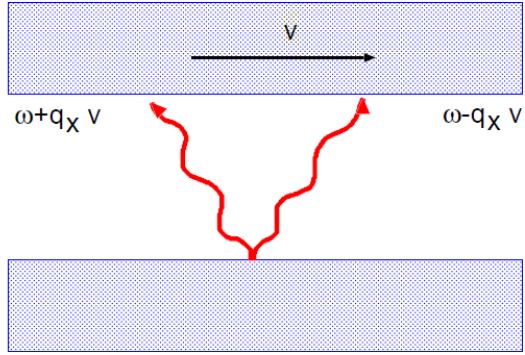}
\caption{\label{Doppler} The electromagnetic waves emitted in the opposite
direction by the body at the bottom will experience  opposite
Doppler shift in the reference frame in which the body at the top
is at rest. Due to the frequency dispersion of the reflection
amplitude these electromagnetic waves will reflect differently
from the surface of the body at the top, which gives rise to
momentum transfer between the bodies. This momentum transfer is
the origin of the van der Waals friction.}
\end{figure}

The origin of the van der Waals friction can be explained by the Doppler effect. Let us consider two flat parallel surfaces, separated by a sufficiently
wide vacuum gap, which prevents electrons from tunneling across
it. If the surfaces are in relative motion (velocity $v$) a frictional stress will act between them. This frictional stress is
related with an asymmetry of the reflection amplitude along the direction of motion; see Fig. \ref{Doppler}. If one body emits radiation, then
in the rest reference frame of the second body these waves are Doppler shifted which will result in different reflection
amplitudes. The same is true for radiation emitted by the second body. The exchange of ``Doppler shifted photons'' is the origin of
van der Waals friction.

From the point of view of the quantum mechanics the van der Waals friction originates from two types of processes. (a) If in the rest reference frame of moving body excitation has a frequency $\omega_{\alpha}(\mathbf{q})$ then in the laboratory reference frame this excitation will have frequency $\omega_{\alpha}(\mathbf{q})-q_xv$. If $\omega_{\alpha}(\mathbf{q})-q_xv<0$ then as a result of such process the photon can be excited with frequency $\omega=q_xv - \omega_{\alpha}(\mathbf{q})$.
 (b) An excitation annihilated in the rest reference frame of
moving body will create photon with frequency $\omega=\omega_{\alpha}(\mathbf{q}) - q_xv$ in the laboratory reference frame. The photons created in processes (a) and (b) can excite excitations in the body which is at rest in the laboratory reference frame that will result in momentum transfer and friction.  Thus in process (a) the excitations are excited in both bodies, in contrast to process (b) for which excitation is annihilated in one body and created in other body. The first process (a) is possible
even at zero temperature, when it is associated with quantum friction. The second process (b) is possible only
at finite temperatures. Quantum and thermal friction are associated with quantum and thermal fluctuations, respectively. The process (a) will dominate at $v>dk_BT/\hbar$ and process (b) will dominate at $v<dk_BT/\hbar$. At small velocities thermal and quantum friction forces depend linearly and cubically on sliding velocity, respectively.

The van der Waals friction is related with exchange between bodies of real photons. This fact was overlooked by Leonhardt \cite{LeonhardtNJP2010}. He assumed that the asymmetry of reflection coefficients, which appears for the moving bodies due to the Doppler effect, can be obtained also for a magneto-electric material where the magneto-electric coupling plays the role of the velocity. If close to  a magneto-electric material another dielectric is placed, between them will act the force, analogous to the friction force, which appears between  bodies moving relative to each other. According to the Leonhardt arguments  if quantum friction exists, it would lead to the acceleration of dielectric, located near magneto-electric material, which would correspond to the unlimited energy source.
Quantum friction thus leads to the paradox that an unlimited amount of useful energy could be extracted from the quantum vacuum. The error of these arguments consists in the fact that there is no exchange of real photons between bodies both of which are at rest at absolute zero temperature because there are no excitations in this case. Thus, in this case the force between bodies  will be equal to zero. However, if between magneto-electric material and dielectric a temperature difference exists,   a force will act between them. Therefore it is possible to imagine  the heat engine, which converts heat into the useful work by means of the force, analogous to the van der Waals friction force. Such device can find  broad applications in MEMS and NEMS.

It is important to  note that only evanescent  waves give contribution to the quantum friction, i.e. it  exists only between the closely spaced bodies. For the body, which is moved in the absolute vacuum, quantum friction (in contrast to the thermal friction) is equal to zero, what is  in agreement  with   the principle of relativity.

\begin{figure}
\includegraphics[width=0.45\textwidth]{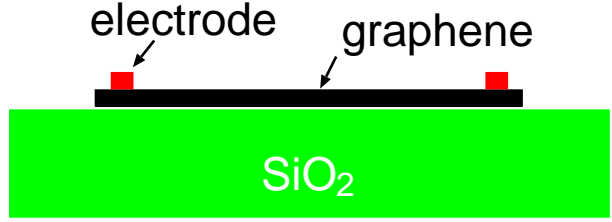}
\caption{\label{transistor} Scheme of the graphene field effect transistor}
\end{figure}

We recently proposed  \cite{VolokitinPRL2011} that  quantum friction can be detected in experiment studying electrical transport phenomena in nonsuspended graphene on amorphous SiO$_2$ substrate (Fig.\ref{transistor}). The electrons, moving in graphene under the action of electric field will experience intrinsic friction due to interaction with the acoustic and optical phonons in graphene and extrinsic friction due to interaction with the optical phonons in located nearby substrate from SiO$_2$. In  high electric fields the electrons move with the high velocities. In this case the main contribution to the friction gives interaction with the optical phonons in graphene and in SiO$_2$. However, the frequency of optical phonons in graphene approximately is four times larger than their frequency in SiO$_2$. Therefore, the main contribution to the friction will give interaction with the optical phonons in SiO$_2$; accordingly, in the high electric fields the electrical conductivity of graphene is determined by the same interaction. It leads to the friction, which we described within the framework of the theories of the van der Waals friction.

\begin{figure}
\includegraphics[width=0.70\textwidth]{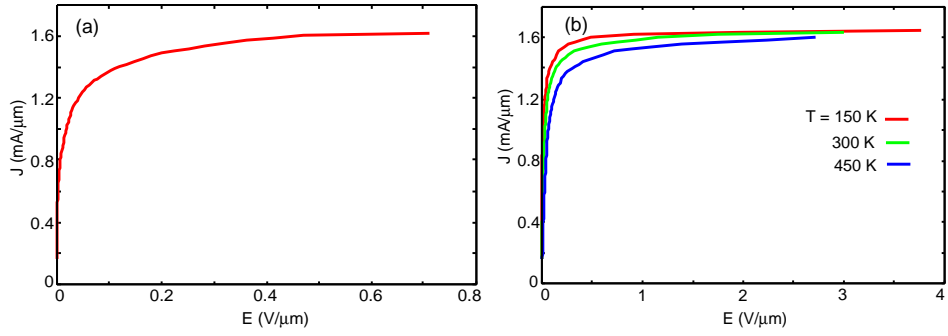}
\caption{\label{Current} The role of the interaction between phonon
polaritons in SiO$_2$ and free carriers in graphene for graphene
field-effect transistor transport. The separation between graphene
and SiO$_2$ is $d=3.5$\AA.  (a) Current density-electric field
dependence at $T=0$ K, $n=10^{12}$cm$^{-12}$. (b) The same as in
(a) but for different temperatures.}
\end{figure}

Figs. \ref{Current}a and \ref{Current}b show the dependence of the current density on the electric field at $n=10^{12}$cm$^{-2}$, and for
different temperatures.  In obtaining these curves we have used that $J=nev$ and $neE=F_x$, where $J$ and $E$ are current density
and electric field, respectively, $F_x$ is the frictional stress acting on electrons (or holes) in graphene.  Note that the current density saturates at $E\sim 0.5-2.0$V/$\ \mu$m, which is in agreement  with
experiment \cite{FreitagNL2009}. The saturation velocity can be extracted from the $I-E$ characteristics using
$J_{sat}=nev_{sat}$, where 1.6 mA/$\mu$m is the saturated current density, and with the charge density concentration
$n=10^{12}$cm$^{-2}$: $v_{sat}\approx 10^6$m/s. Fig. \ref{Current}a was calculated at $T=0$ K. At zero temperature the van der Waals
friction is due to quantum fluctuations of charge density,

According to the theory of the van der Waals friction \cite{VolokitinRMP2007} (see also above discussion), the quantum friction, which exists
even at zero temperature, is determined by the creation of excitations in each of the interacting media, the frequencies of
which are connected by $vq_x=\omega_1 + \omega_2$. The relevant excitations in graphene are the electron-hole pairs whose
frequencies begin from zero, while for SiO$_2$ the frequency of surface phonon polaritons $\omega_{ph} \approx 60$meV ($9\cdot
10^{13}$s$^{-1}$). The characteristic wave vector of graphene is determined by Fermi wave vector $k_F$. Thus the friction force is
strongly enhanced when $v>v_{sat}=\omega_{ph}/k_F \sim 10^6$m/s, in  accordance with numerical calculations. Thus  measurements of
the current density-electric field relation of graphene adsorbed on SiO$_2$ give the possibility to detect quantum friction.

\section{Friction induces an electric field}

\begin{figure}
\includegraphics[width=0.45\textwidth]{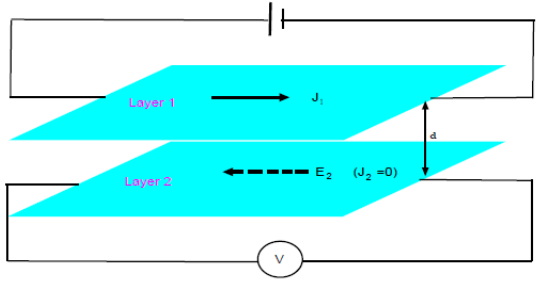}
\caption{\label{Drag} Scheme of experiment for observation of  the drag effect.}
\end{figure}

An alternative method of studying of the van der Waals friction consists in driving an electric current in one metallic layer and
studying of the effect of the frictional drag on the electrons in a second (parallel) metallic layer (Fig. \ref{Drag}). Such experiments were predicted by Pogrebinskii \cite{Pogrebinskii} and Price \cite{Price} and were performed for 2D quantum wells \cite{Gramila,Sivan}. In these
experiments a current  is driven through layer \textbf{1}.  Due to the proximity of the layers, the interlayer interactions will induce a current in layer \textbf{2} due to a friction stress  acting on the electrons in the layer \textbf{2} from layer \textbf{1}. If layer \textbf{2} is an open circuit, an electric field $E_1$ will develop in the layer whose influence cancels the frictional stress
$\sigma$ between the layers.

Similar to 2D-quantum wells in semiconductors, frictional drag experiments can be performed (even more easily) between graphene
sheets. Such experiments can be performed in a vacuum where the contribution from the phonon exchange can be excluded. To exclude
noise (due to presence of dielectric) the frictional drag experiments between quantum wells were performed at very low
temperature ($T\approx 3 \ {\rm K}$)\cite{Gramila}. For suspended graphene sheets there is no such problem and the experiment can be
performed at room temperature. In addition, 2D-quantum wells in semiconductors have very low Fermi energy $\epsilon_F \approx
4.8\times 10^{-3}$eV \cite{Gramila}. Thus electrons in these quantum wells are degenerate only for very low temperatures
$T<T_F=57$ K. For graphene the Fermi energy $\epsilon_F=0.11$eV at $n=10^{12}$cm$^{-2}$, and the electron gas remains degenerate for
$T<1335$ K.

\begin{figure}
\includegraphics[width=0.70\textwidth]{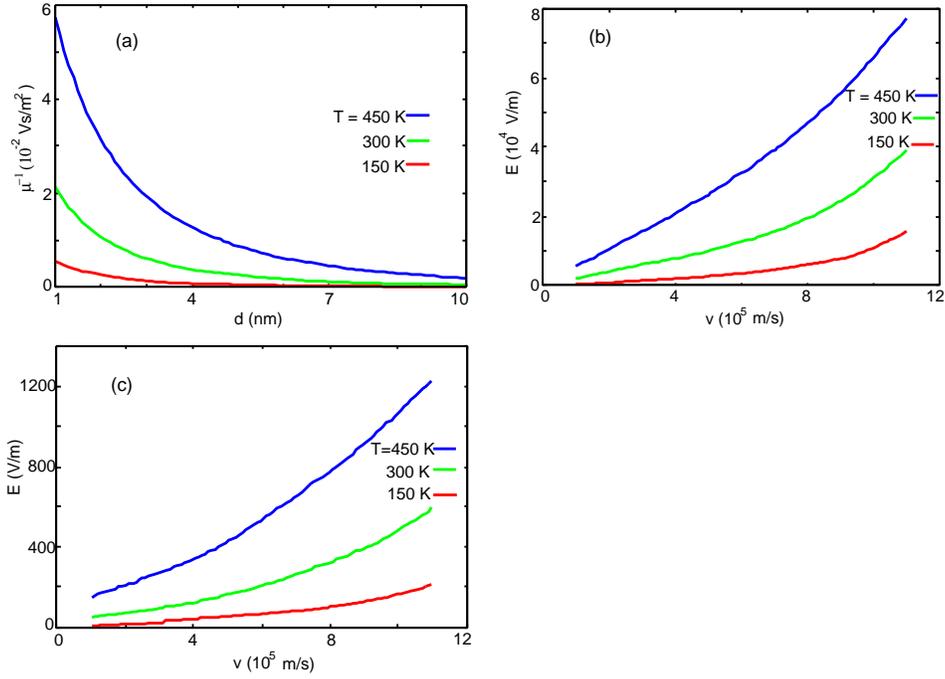}
\caption{\label{DragEffect} Frictional drag between two graphene sheets
at the carrier concentration $n=10^{12}$cm$^{-2}$.  (a) Dependence
of friction coefficient per unit charge, $\mu^{-1}=\Gamma/ne$, on
the separation between graphene sheets $d$. (b) Dependence of
electric field induced in graphene on drift velocity of charge
carriers in other graphene sheet at the layer separation $d=1$ nm.
(c) The same as in (b) but at $d=10$nm.}
\end{figure}

At small velocities the electric field induced by frictional drag depends linearly on the velocity, $E=(\Gamma/ne)v = \mu^{-1}v$,
where $\mu$ is the low-field mobility. Figure \ref{DragEffect}a shows the dependence of the friction coefficient
(per unit charge) $\mu^{-1}$ on the separation $d$ between the sheets. For example, $E=5\times 10^{-4}v$ for $T=300$ K and $d=10$
nm. For a graphene sheet of length $1\ {\rm \mu m}$, and with $v=100$m/s this electric field will induce the voltage $V=10$ nV.
Figures \ref{DragEffect}b  and \ref{DragEffect}c show the induced electric field-velocity relation for high velocity, with $d=1$nm (b) and
$d=10$nm (c). As one can see, the effect of frictional drag between the graphene sheets strongly depends on temperature, i.e. it is determined by the thermal fluctuations.

\section{Breakdown of  Stephan-Bolzmann law  at the nanoscale}

 Transfer of energy between two surfaces separated by vacuum gap is a topic that has fascinated several  generations of researches. If both surfaces
 are at rest then at large separation $d\gg \lambda_T=k_BT/\hbar$ the radiative heat transfer is determined by
the Stefan-Boltzman law, according to which the thermal heat transfer coefficient  $\alpha=4\sigma T^3$. In this limiting case the heat
transfer between two bodies is determined by the propagating electromagnetic waves radiated by the bodies, and does not depend
on the separation $d$. At $T=300$K this law predicts the (very small)  heat transfer coefficient, $\alpha \approx 6$
Wm$^{-2}$K$^{-1}$. However, as was first predicted theoretically by Polder and Van Hove \cite{Polder} in the framework of
stochastic electrodynamics introduced by Rytov \cite{Rytov}, and recently confirmed
experimentally \cite{Chen2009a,Greffet2009}, at short separation $d\ll \lambda_T$, the heat transfer may increase by many orders of
magnitude due to the evanescent electromagnetic waves; this is often referred to as photon tunneling. Particularly strong
enhancement occurs if the surfaces of the bodies can support localized surface modes such as surface plasmon-polaritons,
surface phonon-polaritons, or adsorbate vibrational modes \cite{VolokitinRMP2007}. The findings  have an influential role in new concepts such as thermophotovoltaics, NEMS, heat-assisted magnetic recording, heat-assisted lithography, the design of devices that rely on resonant or coherent heat emission, and nano-antennas.

\begin{figure}
\includegraphics[width=0.45\textwidth]{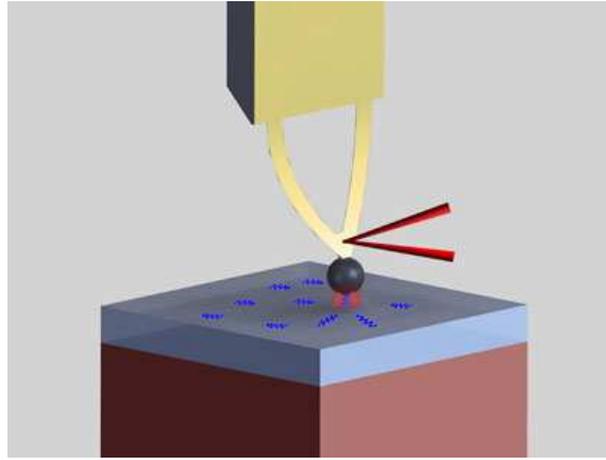}
\caption{\label{HeatExperiment} Schematic diagram of experimental setup for observation of the near-field radiative heat transfer between a silica microsphere and the silica substrate.}
\end{figure}

In two recent experiments \cite{Chen2009a,Greffet2009} the heat transfer coefficient between a heated, flat surface and a micrometric-scale sphere mounted on the bimorph cantilever of an atomic force microscope  was measured over a distance of  30 nm to several micrometers. Heat transfer across the plate-sphere gap causes the  cantilever to bend very slightly, and this is measured by optical fibre interferometry (Fig.\ref{HeatExperiment}). Experiments were restricted to separation wider than 30 nm, because here the imperfections start to distort the measurements of heat transfer. The extreme near-field -- separations less than approximately 10 nm -- may not be accessible using a plate-sphere geometry because it is extremely difficult to fabricate well defined extended planes and high-quality sphere at the nanoscale without unwanted surface spots.   Such an extreme short separation is of great interest to those who design nanoscale device, as modern nanostructures are considerably smaller than 10 nm and are separated in some cases by only a few fraction of a nanometer.  Recently we have shown \cite{VolokitinPRB2011} that  studing the radiative heat transfer at extremely short separation in the plate-plate configuration is also very convenient with the aid of graphene.

Suspended graphene sheet has a roughness $\sim$1 nm (Ref.\cite{Meyer2007}), and measurements of the thermal conductance can
be performed from separation larger than $\sim$ 1 nm.  Another advantage of using
graphene for studying the radiative heat transfer result from the fact that under steady-state condition the heat flow is equal to the heat generated in the graphene by the current density: $S_z=EJ$. This quantity can be accurately obtained from $I-V$ characteristics.  The temperature of graphene can be measured
accurately using Raman scattering spectroscopy \cite{FreitagNL2009}.

\begin{figure}
\includegraphics[width=0.80\textwidth]{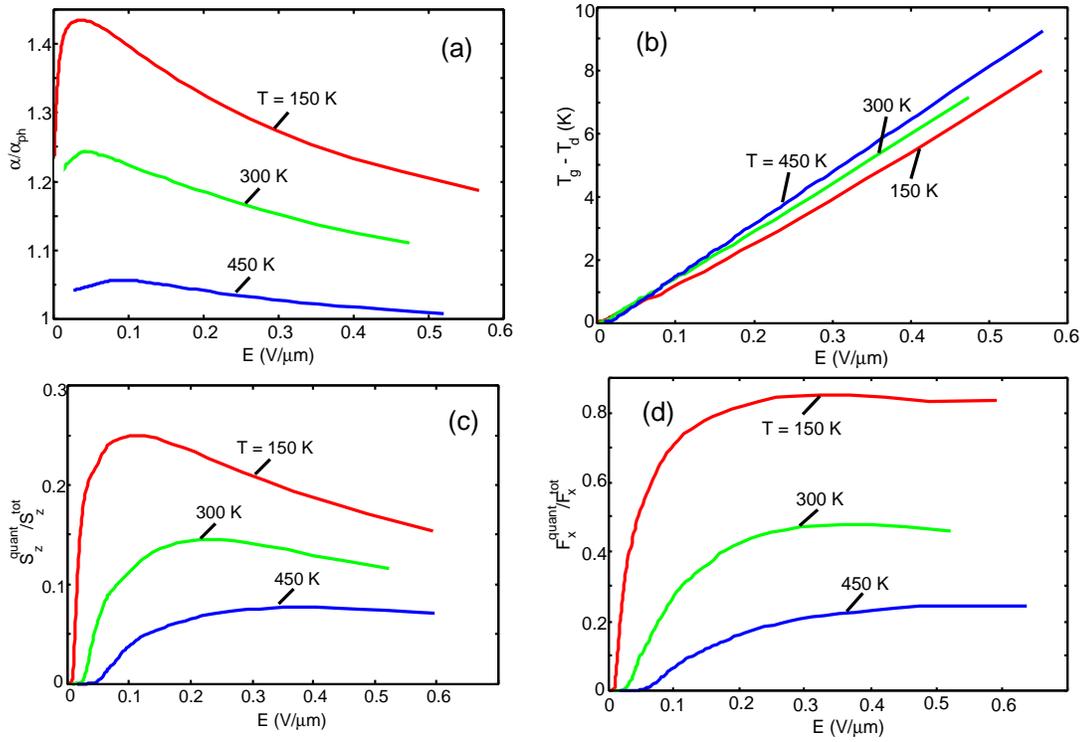}
\caption{\label{HeatTransfer} Radiative energy transfer between graphene   and
SiO$_2$ for $n=10^{16}$m$^{-2}$, $d=0.35$ nm and $\alpha_{ph}=1.0\times
10^8$Wm$^{-2}$K$^{-1}$. (a) The dependence of the ratio between the total energy
transfer coefficient and the phononic heat transfer coefficient, on electric
field.  (b) Dependence of the temperature difference
between graphene and substrate on the electric field.
 (c) Dependence of the ratio between the heat flux only due to quantum fluctuations $S_z^{quant}$ and the
total  energy flux, on the electric field. (d) Dependence of the ratio between the friction force  only due to quantum fluctuations $F_x^{quant}$ and the
total friction force, on the electric field.}
\end{figure}

For nonsuspended graphene besides the radiative mechanism of the heat transfer there is also a mechanism, connected with interaction between the vibrations in the contacting media. With the presence of this phononic mechanism, the  heat generated in graphene by the current density  will be transferred to the substrate by means of the radiative and phononic mechanisms: $EJ = S_z +
\alpha_{pn}(T_g - T_d)$, where $\alpha_{pn}$ is the phononic heat transfer coefficient, $T_g$ and $T_d$ - the graphene and dielectric temperatures, respectively.
Accordingly to theoretical calculations\cite{Persson2010c} and experiment\cite{Chen2009} at room temperature the heat transfer coefficient due to the direct phononic coupling for the interface between graphene and a perfectly smooth (amorphous) SiO$_2$ substrate is $\alpha_{\rm ph} \approx 1.0\times
10^8$Wm$^{-2}$K$^{-1}$.

 A  general theory of the radiative energy transfer between moving bodies, with arbitrary relative velocities, was developed by us in Ref.
\cite{VolokitinPRB2008}. According to this theory there is transfer of energy between moving bodies even at zero temperature
difference, and the heat is generated by the relative motion of quantum and thermal fluctuations. It appears in its most elementary form when the surfaces are at zero Kelvin and the heat is generated  by the relative movement of quantum fluctuations. In  Ref. \cite{VolokitinPRB2011} this theory was
applied to calculate the radiative energy transfer between carriers (moving with the drift velocity $v$) in graphene and the
substrate. The motion of the charge carriers in graphene  relative to the substrate leads to  new effects. In absence of relative motion  the heat transfer between bodies  is determined only by thermal fluctuations. For the moving relative to each other bodies   both thermal and quantum fluctuations  give  contributions to the  radiative heat transfer.

Fig. \ref{HeatTransfer}a shows the ratio of the energy transfer coefficient to the phononic heat transfer coefficient for $d=0.35$ nm and $n=10^{16}$ m$^{-2}$.
 For low and intermediate field this ratio is larger than unity what means that in this region the near-fields
 radiative energy transfer gives additional significant contribution to the heat transfer due to direct phononic coupling.
For nonsuspended graphene on SiO$_2$ the energy and heat transfer are very effective thus the temperature difference does not rise high even
for high electric field when  saturation in $I-E$ characteristic begins \cite{FreitagNL2009} (see Fig. \ref{HeatTransfer}b). The radiative
heat transfer between bodies at rest is determined only by thermal fluctuations, in contrast to the radiative energy transfer between
moving bodies which is determined by both thermal and quantum fluctuations.  Fig. \ref{HeatTransfer}c shows that quantum fluctuations can give
significant contribution to the total energy transfer for low temperatures and large electric field (high drift velocity). Similarly,
in the (elecric current) saturation region quantum fluctuations give significant contribution to the total friction force which is determined, as discussed
above, by the sum of the extrinsic and intrinsic friction forces (see  Fig. \ref{HeatTransfer}d) . The extrinsic friction force has contributions
from both thermal and quantum  fluctuations.

At large separation between graphene and the substrate ($d>1$ nm) the heat transfer between them is determined only by radiative mechanism. At $d \sim$ 5 nm and $T=300$ K the energy transfer coefficient, due to the near-field radiative energy transfer, is
$\sim 10^4$Wm$^{-2}$K$^{-1}$, which is $\sim 3$ orders of magnitude larger than the radiative heat transfer coefficient of the black-body radiation. In
comparison, the near-field radiative heat transfer coefficient  in SiO$_{2}$-SiO$_{2}$ system for the plate-plate configuration, when
extracted from experimental data \cite{Chen2009a} for the plate-sphere configuration, is $\sim $
2230Wm$^{-2}$K$^{-1}$ at a $\sim$30 nm gap. For this system the radiative heat transfer coefficient depends on separation as
$1/d^2$; thus $\alpha \sim 10^5$Wm$^{-2}$K$^{-1}$ at $d\sim 5$ nm what is one order of magnitude larger than for the
graphene-SiO$_2$ system in the same configuration. However, the sphere has a characteristic roughness of $\sim$ 40 nm, and the
experiments \cite{Chen2009a,Greffet2009} were restricted to separation wider than 30 nm (at smaller separation the
imperfections affect the measured heat transfer). Thus the extreme near-field-separation, with $d$ less than approximately 10 nm, may
not be accessible using a plate-sphere geometry. A suspended graphene sheet has a roughness $\sim$1 nm, and measurements of the heat transfer coefficient can be performed from separation larger than $\sim$ 1
nm.  This range is of great interest for the
design of nanoscale devices, as modern nanostructures are considerably smaller than 10 nm and are separated in some cases by
only a few Angstroms.

\section{Conclusion}

Quantum friction as superconductivity and superfluidity, is the macroscopic phenomenon, which nature is determined by quantum laws. This determines its fundamental significance, but quantum friction can  be also technologically important. For example, characteristics  of the graphene field-effect transistor  can be determined in many respects by quantum friction. Accordingly to the concept of nanotechnologists, the graphene field-effect transistor should replace in the next 10 years silicon transistor and provide further progress in nanoelectronics. Many others micro- and the nano-electromechanical systems, which promise new applications in the sensors and the information technologies, can depend on the presence of  quantum friction. Furthermore, quantum friction determines the limit, to which the friction force can be reduced  and, consequently, also the force fluctuations, since, according to the relationship established by Einstein, the friction and fluctuations are connected with each other. According to this relationship, the random force that makes a small particle jitter would also cause friction if the particle is dragged through the medium. From the other side the force fluctuations are important for ultrasensitive force detection experiments.
For example, the detection of single spins by magnetic resonance force microscopy \cite{Rugar}, which has been proposed for
three-dimensional atomic imaging \cite{Sidles} and quantum computation \cite{Berman}, will require force fluctuations (and
consequently the friction) to be reduced to unprecedented levels. In addition, the search for quantum gravitation effects at short
length scale \cite{Arkani}, and future measurements of the Casimir  forces \cite {Mohideen}, may eventually be limited by non-contact friction effects, one mechanism of which is determined by the van der Waals friction with its limiting case --  quantum friction.  For these applications better understanding of non-contact friction is only the first step. In the future it will be necessary to learn how to reduce it or, in other words, how to `lubricate'  vacuum. Intriguing idea - creation of heat engine based, as it was discussed above, on  ideas about nature of the van der Waals friction. In such engine the temperature difference between magneto-electrical material and closely spaced dielectric will induce lateral forces between bodies. This effect can find broad application in NEMS.

\vskip 0.5cm
\textbf{Acknowledgment}

The author acknowledges financial support from  ESF within activity
``New Trends and Applications of the Casimir Effect''.

\end{document}